\documentclass[a4paper,12pt]{article}

\usepackage{amsfonts}
\usepackage{amssymb}
\usepackage{amsmath}

\usepackage{cite,bm,comment}
\usepackage{epsf}
\def\eqref#1{(\ref{#1})}
\def\bu{$\bullet$}
\def\beq{\begin{equation}}
\def\eqn#1{\beq\label{#1}}
\def\eeq{\end{equation}}
\def\ee{\end{equation}}

\def\dia{~~$\diamondsuit$}

\def\bea{\begin{eqnarray}}
\def\eqnn#1{\begin{eqnarray}\label{#1}}
\def\eea{\end{eqnarray}}

\def\md{\medskip}

\newcommand{\eqna}[1]{\begin{subequations} \label{#1}
\begin{eqnarray}}
\def\eena{\end{eqnarray}
\end{subequations}}

\def\kc{K_\chi^\t}  
\def\htt{\hat{T}}

\def\nt{\noindent}

\def\rank{{\rm rank}}
\def\hc{\hat{C}}

\def\nd{\end{document}}
 
\def\ha{{\textstyle{1\over2}}}
\def\bbc{\mathbb{C}}
\def\bbr{\mathbb{R}}
\def\hp{\hat{\varphi}}

\def\vf{\varphi}
\def\D{\Delta} 
 \def\l{\lambda} 
\def\a{\alpha} \def\b{\beta} \def\s{{\sigma}}

 \def\rra{\longrightarrow}

\def\nn{\nonumber}

\def\s#1{{\mathfrak s}_{#1}}

 \def\cg{{\cal G}}

\def\np{\newpage}

\def\ra{\rightarrow}

\def\ca{{\cal A}}  \def\cc{{\cal C}}
  
\def\cg{{\cal G}} \def\ch{{\cal H}} 
 \def\ck{{\cal K}} 
\def\cm{{\cal M}} \def\cn{{\cal N}} 
  
\def\cs{{\cal S}}

\def\tcn{\widetilde{{\cal N}}}
\def\tN{\widetilde{N}}
\def\tn{\widetilde{n}}

\def\t{\tau}

\def\tD{{\tilde{D}}}
\font\tfont=cmbx12 scaled\magstep1 
\def\tih{{C_\chi}}  
\def\va{{\vert a\vert}}
\def\vak{{\vert a_k\vert}}

\def\hd{{D}}
\def\s{\sigma}

\begin{document}

\begin{center}

\centerline{{\tfont Tenfold Way for Holography : AdS/CFT and Beyond }}

\vskip 1.5cm

{\bf V.K. Dobrev}

 \vskip 5mm

  Institute for Nuclear Research and Nuclear Energy,\\ Bulgarian
Academy of Sciences,\\ 72 Tsarigradsko Chaussee,  1784 Sofia,
Bulgaria

 \end{center}

\vskip 1.5cm

 \centerline{{\bf Abstract}}

The main purpose of the present paper is to
   lay the foundations of generalizing the AdS/CFT (holography) idea   beyond the conformal setting.
The main tool is to find suitable realizations of the bulk and boundary via group theory.  We use all ten families of classical real semisimple Lie groups $G$
and Lie algebras $\cg$. For this are used
several group and algebra decompositions: the global Iwasawa decomposition   and the local Bruhat and Sekiguchi-like
decomposititions. The same analysis is applied to the exceptional real semisimple Lie algebras.

\vskip 1.5cm


\np

\section{Introduction}

For the last twenty years due to the remarkable proposal of
\cite{Malda} the AdS/CFT correspondence is a dominant subject in
string theory and conformal field theory. Actually the possible
relation of field theory on anti de Sitter space to conformal field
theory on boundary Minkowski space-time was studied also before,
cf., e.g., \cite{FlFr,AFFS,Fro,BrFr,NiSe,FeFr}. The proposal of
\cite{Malda} was further elaborated in \cite{GKP} and \cite{Wita}.
After that  there was an explosion of related research which
continues also currently.

Let us recall that the AdS/CFT correspondence
has 2 ingredients \cite{Malda,GKP,Wita}:
1. the holography principle, which is very
old, and  means the reconstruction of some objects in the bulk (that
may be classical or quantum) from some objects on the boundary;
 2. the reconstruction of quantum objects, like 2-point functions on the
boundary, from appropriate actions on the bulk.

Our focus here is on the first ingredient. For further reference we note that in most applications to physics the boundary
is interpreted as the surface of the bulk, (see e.g.  \cite{Malda,GKP,Wita,Busso}), thus their dimensions differ by one  and we shall
follow this. 

We note that   the first explicit presentation of
the holography principle was realized in the Euclidean case, \cite{Wita}, i.e., for the Euclidean conformal group ~$G_E = SO_0(d+1,1)$.
In this case the bulk space ~$\cs_0$~ is isomorphic  to the factor-space:
\eqn{bulke}  \cs_0 ~=~ G_E/K = SO_0(d+1,1)/SO(d+1) \ee
where ~$K=SO(d+1)$~ is the maximal compact subgroup of ~$G_E$.
It is important that in this case we use the so-called Iwasawa decomposition ~$G_E=KA_0N_0$ (explained mathematically below, see also \cite{War,Dob1,Dobpar}),
where the subgroups ~$A_0$, ~$N_0$, are important from the physics point of view, namely
 ~$A_0$~ is the subgroup of dilatations, ~$N_0$~ is the subgroup of Euclidean  translations (isomorphic to the subgroup of special conformal transformations,
and also to the $d$-dimensional Euclidean space, $R^d$). Note that the dimension ~$r_0$~ of $A_0$ is called the ~{\it split rank} of $G$.

Thus, we have for the bulk: ~$\cs_0 \cong A_0N_0$, while the boundary is isomorphic to ~$N_0$ \cite{Dobads}. Note that this realization of
the bulk is very suitable for the applications since the parametrization of $A_0$ provides and easy limit from the bulk to the boundary. 

As historical {\bf Remark} we mention that the problem is related to the
construction of the discrete series of unitary representations in
\cite{Hot,Schm}, which was then applied in \cite{DMPPT} for the Euclidean conformal group $SO_0(4,1)$ (also for $SO_0(2,1)$).
The approach applied to the
 Euclidean case $G_E$ in \cite{Dobads} is different. We should mention that the nonrelativistic Schr\"odinger
   case was considered in \cite{AiDo} using the representation theory developed in \cite{ADDMS} (for an invite review see \cite{DobNR}).

The next task was to  show the holography principle
 for Minkowski space-time, i.e., for the conformal
group ~$G_M=SO_0(d,2)$.  Initially, this was done relying on Wick rotations of the final results, cf., e.g. \cite{Wita},
 (see also some other early papers
\cite{HeSf,MuVi,FMMR,LiTs,MeTs,CKPF,LMRS,AruFro,Akh,Imamura,Corl,Vol,DHoFre,BCFM,ChaSch,Bon,BilChu,GuMiZa,Yi}). For some
more recent papers and reviews see, e.g.,
\cite{Mald,MalWit,Beisert,Myers,Arut,DHKS,KMMZ,Staud,ESZ,BKKS,CPPR,AHS,McGre,Gaber,Polchinski,Hubeny,Gromov,Harlow,Liu}.

 Of course, it is  desirable to apply group theory tools directly to ~$G_M$. The problem here is that the
 Iwasawa decomposition ~$SO_0(d,2) =  K_M A_M N_M$, is not suitable for the physics applications since ~$A_M$~ is two-dimensional, (thus, there is
 no natural parametrization of the limit from the bulk to the boundary), 
 ~$K_M=SO(d)\times SO(2)$,
  and  ~$N_M$~ has dimension $2d-2$ (bigger than $d$ for $d>2$). Fortunately, there is a suitable
 group decomposition, called Bruhat decomposition, which has the necessary group-theory properties and useful physical interpretation.
 Namely, it is:
 \eqn{bruh}  G_M=SO_0(d,2) ~=_{loc}~ \tN M A N \ee
 where $N,\tN$~ are isomorphic d-dimensional spaces, also isomorphic to d-dimensional Minkowski space-time ~{\bf M},
 ~$A$~ represents the one-dimensional dilatations, ~$M=SO_0(d-1,1)$~ are the Lorentz transformations of  ~{\bf M}.
 It is easy to see that in this setting the role of bulk space is played by ~$\cs_M = AN$ which can be obtained by Wick rotation from
 the Euclidean  $\cs_0 \cong A_0N_0$, while the boundary is Wick rotation from $N_0$ to $N$.  This decomposition was used in our setting
 for $d=3$ in \cite{AizDob}.\\
 {}[On the technical side : the designation ~$_{loc}$ above means that the subgroup $\tN M AN$ is an open dense set of $G_M$.]

\md\nt {\it Remark:} ~~One may ask why not use the Bruhat decomposition also in the split rank one cases. Indeed, this is possible and is a matter of choice.
Our choice is motivated by our experience that calculations for split rank one case are easier with the Iwasawa decomposition.\dia  

 One more important ingredient in the present paper is the fact that the space ~$\cs_M$~ may arise also by using another decomposition
 first introduced for some groups by Sekiguchi \cite{Seki}, though with no relation to our setting. For the conformal group
 this decomposition is:
 \eqn{sekigg} SO_0(d,2) ~=_{loc}~  H A N, ~~~H= SO_0(d,1)  \ee
 Note that this decomposition may be obtained via Wick rotation from the Iwasawa decomposition: ~$SO_0(d+1,1) \ra SO_0(d,2)$,
 ~$SO(d+1) \ra SO_0(d,1)$. We initiated the use of this decomposition in our setting in \cite{DoMo2}.
 There it was shown that although the bulk space ~$\cs_M = AN$~
 obtained via ~$SO_0(d,2)/\tN M$ ~is isomorphic to the one obtained via ~$SO_0(d,2)/H$, the actual parametrizations in terms of
 the groups elements of ~$SO_0(d,2)$~ do not coincide. Furthermore, the parametrization obtained from \eqref{sekigg} is simpler
 which is important for the applications.

 This decomposition is not universal as it does not exist for all real semisimple Lie groups. In \cite{Seki} the Sekiguchi decomposition
 was defined for ~$SO_0(p,q)$, $SU(p,q)$, $Sp(p,q)$.  Following the idea of \cite{Seki} we define Sekiguchi-like\footnote{We reserve the term 'Sekiguchi decomposition'
 for the cases contained in \cite{Seki}.} decomposition
 for a real semisimple Lie group $G$ as follows.
Let there be a Bruhat decomposition
\eqn{bruhh} G ~=_{loc}~  \tN M A N,\ee
such that there exists a subgroup ~ $H$~ of ~$G$,
so that the Sekiguchi-like decomposition
\eqn{sekiggg}  G ~=_{loc}~  H A N  \ee
exists with the same subgroups $AN$. Following the case of the conformal group ~$SO_0(d,2)$~ we expect that in the cases when \eqref{sekiggg} exists
it will be simpler to use than the Bruhat decomposition. Below we give many examples of Sekiguchi-like decomposition beyond the list of \cite{Seki}.

\md

   Now we can state the ~{\bf main purpose of the present paper}: to
   lay the foundations of generalizing the holography (AdS/CFT) idea beyond the conformal setting.

What we do is to consider the real Lie groups and to explicate for each of them
 their  subgroups mentioned above in the conformal setting.

The words ~{\bf tenfold way}~ in the title refer to the fact that there are ten classical real Lie groups when we include the complex
classical groups  considered over the reals. We should note that the tenfold way has many other manifestations
in mathematics and physics which are well described on John Baez' web-page \cite{Baez} (see especially \cite{AlZi,RSFL,LMC}
for applications to solid state physics).  Our approach differs from others since we use
~{\it noncompact} Lie groups which is essential in the holography applications.


Let us {\bf summarize} what we got as guidelines from the above. In the cases of $G$ of split rank $r_0=1$ most suitable is the Iwasawa
decomposition $G=KA_0N_0$ in which case the bulk space is $\cs_0  = A_0N_0$ and the boundary space is $N_0$. In the cases of split rank $r_0>1$
we would use the Sekiguchi-like decomposition \eqref{sekiggg} when it exists, otherwise we shall use the Bruhat decomposition \eqref{bruhh}. In the
latter cases the bulk space is $\cs  = AN$ and the boundary space is $N$.

The paper is organized as follows. In Section 2 we introduce more systematically the necessary group theory prerequisites.
In Section 3 we consider the bulk-boundary correspondence for the special case of groups of split rank 1.
In Section 4 we consider the bulk-boundary correspondence for the cases of split rank $>$1.
There are five Tables containing our analysis and results of the relevant group-theory data. The tables are placed at the end of the paper in order
not to interrupt the exposition.

\section{Preliminaries}

\subsection{Lie group and algebra decompositions}

We need some well-known preliminaries to set up our notation and
conventions (cf. e.g., \cite{War}, see also \cite{Dob1,Dobpar}). Let $G$ be a noncompact semisimple Lie group.  Let $K$ denote a
maximal compact subgroup of $G$. Then we have  the global {\it Iwasawa}
decomposition:
\eqn{iwas} G ~=~ KA_0N_0,\ee
 where ~$A_0$~ is abelian simply connected, a
vector subgroup of ~$G$, ~$N_0$~ is a nilpotent simply connected
subgroup of ~$G$~ preserved by the action of ~$A_0$. This decomposition is called global since
every element $g\in G$ may be represented as the product of three elements of the corresponding
subgroups, namely, $g= ka_0n_0$, ~$k\in K, ~a_0\in A_0, ~n_0\in N_0\,$.\footnote{Note that the order
of the three factors of the Iwasawa decomposition may vary, then of course the elements
representing the subgroups changes.} Note that there is another nilpotent subgroup
~$\tN_0$, which is isomorphic to $N_0$, and that there is analogous Iwasawa decomposition ~$G= KA_0\tN_0$.

 Further, let $M_0$
be the centralizer of $A_0$ in $K$. Then the subgroup ~$P_0 ~=$ \\ $= M_0 A_0
N_0$~ is a {\it minimal parabolic subgroup} of $G$.

Further, let $M'_0\supset M_0$ be the normalizer of $A_0$ in $K$. The finite group ~$W= W(G,A_0) = M'_0/M_0$~
is called the Weyl group for the pair $(G,A_0)$, or a restricted Weyl group. It has two elements $W=\{ 1,w\}$. The nilpotent subgroups
~$N_0$ and $\tN_0$ are conjugate under the Weyl transformation: ~$wNw^{-1} = \tN_0$.

A parabolic subgroup $P
~=~ M A N$ is any subgroup of $G$
which contains a minimal parabolic subgroup. The number of non-conjugate
parabolic subgroups is ~$2^{r_0}-1$, where $r_0=\rank\,A_0$, called the {\it split rank} of $G$.\footnote{In some expositions
authors are counting as a parabolic subgroup also the group $G$, then the number of non-conjugate
parabolic subgroups would be ~$2^{r_0}$.}\\

Note that in general $M$ is a reductive Lie group with structure:
~$M=M_d M_s M_a$, where $M_d$ is a finite group, $M_s$ is a
semisimple Lie group, $M_a$ is an abelian Lie group central in $M$.\\
For further use we note that the Abelian group $A$  may be used as the product
of one-dimensional subgroups:
\eqn{zar} A=A_1 \cdots A_r, ~~~ a\in A, ~~a=a_1\cdots a_r, ~~a_k\in A_k \ee

Another important decomposition is the local {\it Bruhat} decomposition which exists for every parabolic subgroup $P=MAN$:
\eqn{bruh2} G ~=|_{\rm loc}~ \tN MAN, \ee
where ~$\tN$~ is is a nilpotent simply connected
subgroup of ~$G$~ preserved by the action of ~$A$, conjugate to $N$. (This decomposition is called local
since there is a subset of elements of $G$ of lower dimension which can not
  be represented as the product of four elements of the corresponding subgroups.)

An important class of parabolic subgroups are the ~{\it maximal parabolic subgroups}~ which
are defined by the property that  ~$r=\rank\,A=1$.\footnote{Note that when the split rank $r_0=1$, then the minimal
parabolic subgroup is also maximal.} In that case the restricted Weyl group ~$W(G,A)$~ has also two elements.

We need also the corresponding Lie algebras. Thus, ~$\cg,\ck,\ca_0,\cm_0,\cn_0, \tcn_0$,\\
 $\ca,\cm,\cn, \tcn$,
denote the
Lie algebras of ~$G,K,A_0, M_0, N_0,\tN_0$, $A,M,N,\tN$, resp. We also have the Lie algebra  versions of the Iwasawa decomposition:
\eqn{iwalg} \cg ~=~ \ck \oplus \ca_0 \oplus \cn_0 \ee
and the Bruhat decompositions:
\eqn{brul} \cg ~=~ \tcn \oplus \cm \oplus \ca \oplus \cn   \ee

\subsection{Elementary representations}
\label{ERS}

We start with the most general representations
 called (in the representation theory of semisimple Lie groups)
generalized principal series representations  (cf. \cite{Kn}) (see also \cite{HC}).
In \cite{DMPPT,DP,Dob}
they were called ~{\it elementary representations} (ERs).
They are obtained by induction from parabolic subgroups $P=MAN$. The induction is from
finite-dimensional (nonunitary in general) irreps of $M$, from arbitrary (non-unitary) characters of $A$,
and trivially from $N$. There are several realizations of these
representations. We give now the so-called ~{\it noncompact picture}~
of the ERs - it is the one most often  used in physics.

The representation space of these induced representations
consists of smooth functions on ~$\tN$~ with values in
the corresponding finite-dimensional representation space of ~$M$, i.e.:
\eqn{funn} C_\chi ~=~ \{ f \in C^\infty(\tN,V_\mu)\} \ee
where  $\chi\, =\, [\mu,\nu]$,$\,$ ~$\nu$~ is a character function on $A$,
~$\nu(a) = \nu_1(a_1)\cdots \nu_r(a_r)$,
~$\mu\,$ is a  irrep of $M$,$\,$
$V_\mu\,$ is the finite-dimensional representation space of $\mu$.
\footnote{In addition, these functions have special
asymptotic expansion as suitable parameter(s) on $\tN$ tend to $\infty$.}
The representation ~$T^\chi$~ acts in $\tih$ by:
\eqn{lart}(T^\chi(g) f) (\tn) ~=~ \prod_{k=1}^r \vak^{-\nu_k(a_k)} \cdot  \hd^\mu(m)\,
f ( \tn') \ee
where the nonglobal Bruhat decomposition $g=\tn m a n$ is
used:
\eqn{nnam}   g^{-1}\tn ~=~ \tn'  m^{-1}{a}^{-1} n^{-1}  ~, \quad
g\in G , \, \tn, \tn' \in \tN , \, m\in M,\, a\in A,\, n\in N ,\ee
$\vak$ is a suitable positive function on $A_k$,
$\hd^\mu(m)$ is the representation matrix of $\mu$ in $V_\mu\,$.
\footnote{For the cases with measure zero for which ~$ g^{-1}\tn_x$~
does not have a Bruhat decomposition of the form $\tn man$
the action is defined
separately, and the passage from \eqref{lart} to these special
cases is ensured to be smooth by the asymptotic
properties mentioned above.}
\footnote{The representation space $C_\chi$ can be
thought of as the space of smooth sections of the homogeneous vector
bundle (called also vector ~$G$-bundle) with base space ~$G/P$~ and
fibre ~$V_\l\,$, (which is an associated bundle to the principal
~$P$-bundle with total space ~$G$). Actually, we do not need this
description, but following \cite{Dob}  we replace the above
homogeneous vector bundle  with a line bundle again with base space ~$G/P$. The
resulting functions ~$\hp$~ can be thought of as smooth sections of this line
bundle.}

The importance of the elementary representations comes also from the
remarkable result of Langlands-Knapp-Zuckerman
\cite{Lan,KnZu} stating that every irreducible admissible representation of a
real connected semisimple Lie group $G$ with finite centre is
equivalent to a subrepresentation of an elementary representation
of $G$.\footnote{{\it Subrepresentations}~
are irreducible representations realized on
invariant subspaces  of the ER spaces
(in particular, the irreducible ERs themselves).
The admissibility condition is fulfilled in the
physically interesting examples.}
To obtain a subrepresentation of a topologically reducible ER
one has to solve certain invariant differential equations,
cf. \cite{DMPPT,DP,Dob}.

Finally, we recall that Casimir operators ~$\cc_i$ of $\cg$  have constant values on the ERs:
\eqn{cas} \cc_i(\{ X\})\, \vf(x) ~=~ \chi_i(\mu,\nu)\, \vf(x) ~, \qquad
i=1,\dots,\rank \,G   \ee
where ~$\{X\}$~ denotes symbolically the generators
of the Lie algebra ~$\cg$~ of ~$G$.

\subsection{Bulk representations via Iwasawa decomposition}
\label{BRID}

In the previous subsection we discussed representations
on ~$\tN$~ induced from the parabolic subgroup ~$MAN$~
which is natural since the   subgroup ~$\tN$~ is locally
isomorphic to the factor
space ~$G/MAN$ (via the Bruhat decomposition).
Similarly, it is natural to discuss
representations on  the bulk  space ~$\cs_0\cong \tN_0 A_0$~
which are induced from the maximal compact subgroup ~$K$~
since the solvable group ~$\tN_0 A_0$~ is  isomorphic to the factor space ~$G/K$
(via the Iwasawa decomposition in the version $G=\tN_0 A_0 K$). Namely, we consider
the representation space:
\eqn{funk} \hc_\t   ~=~ \{ \phi \in
C^\infty(\cs_0\,,U_\t) \} \ee
where  $\t\,$ is an arbitrary unitary irrep of $K$,
$U_\t\,$ is the finite-dimensional representation space of $\t$,
with representation action:
\eqn{lartu}
(\htt^\t(g)\phi) (\tn a) ~=~ \tD^\t (k)\, \phi ( \tn' a') \ee
where the Iwasawa decomposition is used:
\eqn{nak}   g^{-1}\tn a ~=~ \tn' a' k^{-1}  ~, \quad
g\in G ,\, k\in K, \, \tn, \tn' \in \tN_0 , \, a, a' \in A_0 \ee
and $\tD^\t(k)$ is the representation matrix of $\t$ in $U_\t\,$.
However, unlike the ERs, these representations are reducible,
and to single out an  irrep equivalent, say, a
subrepresentation of an ER, one has to look for solutions
of the eigenvalue problem related to the Casimir operators \cite{DMPPT},\cite{Dob1}.

\subsection{Bulk representations via Bruhat decomposition}
\label{BRBD}

In  subsection \ref{ERS} we discussed representations
on ~$\tN$~ induced from the parabolic subgroup ~$P=MAN$~
which is natural since the   subgroup ~$\tN$~ is locally
isomorphic to the factor
space ~$G/MAN$ (via the Bruhat decomposition).
Now we shall   discuss
representations  on  the bulk  space ~$\cs\cong \tN A$~
which are induced from the  parabolic subgroup ~$P$~ similarly to \eqref{funn}
 \eqn{funnn} \check{C}_\chi ~=~ \{ f \in C^\infty(\tN A,V_\mu)\} \ee
   The representation ~$\check{T}^\chi$~ acts in $\check{C}_\chi$ by:
\eqn{lartz}(\check{T}^\chi(g) f) (\tn a) ~=~ \prod_{k=1}^r \vak^{-\nu_k(a'_k)} \cdot  \hd^\mu(m)\,
f ( \tn'a') \ee
where the   Bruhat decomposition  is used:
\eqn{nnamz}   g^{-1}\tn a ~=~ \tn' a' m^{-1}  n^{-1}  \ee

\subsection{Bulk representations via Sekiguchi-like decomposition}

These representations are introduced similarly to those using the Iwasawa decomposition. Namely, we consider
the representation space:
\eqn{funh} \hc_\s   ~=~ \{ \phi \in
C^\infty(\cs\,,W_\s) \}, ~~~\cs=\tN A, \ee
where  $\s\,$ is a finite-dimensional irrep of $H$,
$W_\s\,$ is the representation space of $\s$,
with representation action:
\eqn{lartuh}
(\htt^\s(g)\phi) (\tn a) ~=~ \tD^\s (h)\, \phi ( \tn' a') \ee
where the Sekiguchi-like decomposition is used:
\eqn{nah}   g^{-1}\tn a ~=~ \tn' a' h^{-1}  ~, \quad
g\in G ,\, h\in H, \, \tn, \tn' \in \tN , \, a, a' \in A \ee
and $\tD^\s(h)$ is the representation matrix of $\s$ in $W_\s\,$.

\subsection{Table of the ten classical real Lie groups}

Before proceedings further we present a table of the classical real Lie groups in Table 1 containing our analysis of the relevant for our purposes data. (Tables are given
at the end of the paper in order not to interrupt the exposition.)\\
Note that we have included the classical complex Lie groups but considered as real - these
are denoted as types ~$A,BD,C$. Note also that for split real forms the subgroup $M_0$ is trivial and this is designated
by the unit element {\bf e}.

\section{The split rank one case}

We start with the cases of split rank 1. This is natural since this class includes the very important for applications  Euclidean conformal group
 ~$SO_0(p,1)$. Furthermore, this was the first explicit AdS/CFT case considered in \cite{Wita} (see also
  \cite{DMPPT} for ~$SO_0(4,1)$ and $SO_0(2,1)$).

  Our results on the structure of the real Lie algebras of split rank 1 are given in Table 2. Note that most of these cases are of low dimension
 and they are conjugate to special cases of $SO_0(p,1)$ for $p=2,3,5$. For completeness besides the classical cases we have included also
 the only exceptional real Lie algebra of split rank 1 ~: ~$F_{4(-20)}$.

As stated in the introduction we are interested in the group-theoretic aspect of the AdS/CFT correspondence. More precisely, we
consider the relation between the representations on the bulk space ~$\cs_0\cong \tN_0 A_0$~  and the elementary representations on the boundary space ~$\tN_0$.
In the case of $SO_0(p,1)$ these operators for low dimensional representations of $K,M_0$ were given in \cite{Wita}, while the general case was given
in \cite{Dobads}. We recall briefly the main results, introducing some additional notation. We shall use the following
group decomposition for every $k\in K$~:
\eqn{dkk} k ~=~ m(k) k_c\ee
where ~$m(k)$ parametrizes the subgroup $M_0$, while ~$k_c$~ parametrizes
the coset ~$K/M_0$, thus, \eqref{dkk} represents the decomposition of
~$K$~ into its subgroup ~$M_0$~ and the coset ~$K/M_0$~: ~
~$K ~\cong ~ M_0~K/M_0$.  We shall use also the relation ~$K/M_0 \cong \tN_0$~ following from:
\eqn{kanman} G=KA_0N_0 \cong \tN_0 M_0A_0N_0 ~\Rightarrow ~K\cong \tN_0 M_0 ~\Rightarrow ~ K/M_0 \cong \tN_0 \ee
Now we can state the ~{\it Bulk - boundary intertwining relations}:\\
Theorem 1 : \cite{Dobads}~~
1.  {\it Bulk-to-boundary intertwining relation : Let us define the operator:
\eqn{aaa}  L_\chi^\t ~:~  \hc_\t   ~\rra ~C_\chi \, , \ee
with the following action:
\eqn{aab} (L_\chi^\t \phi ) (\tn) ~=~ \lim_{\va\to 0}\
\va^{-\D}\ \Pi^\t_\mu\ \phi (\tn a) \ee
where ~$\D=\nu(a)$, ~$\Pi^\t_\mu$~ is the standard projection operator from the
~$K$-representation space ~$U_\t$~ to the ~$M$-representation space
~$V_\mu\,$, which acts in the following way on the
$K$-representation matrices:
\eqn{prj} \Pi^\t_\mu\ \tD^\t (k) ~=~ \hd^\mu (m(k))\   \Pi^\t_\mu \
\tD^\t (k_c) \ee
where we have used \eqref{dkk}.
Then ~$L_\chi^\t$~ is an intertwining operator, i.e.:
\eqn{aac} L_\chi^\t   \circ  \htt^\t (g) ~=~
 T^\chi (g) \circ L_\chi^\t ~, \quad \forall g\in G ~.\ee
In addition, in \eqref{aab}  the operator ~$\Pi^\t_\mu$~
acts in the following truncated way:}
\eqn{prja} \Pi^\t_\mu\ \tD^\t (k) ~=~ \hd^\mu (m(k))\   \Pi^\t_\mu \ee
2. ~ {\it Boundary-to-bulk intertwining relation :
  The operator which is generically inverse to ~$L_\chi^\t\,$~ and which
   restores a function on de Sitter space $\cs_0$ from its boundary
value is given as   follows:
\eqn{inv} {\tilde L}_\chi^\t
~:~ C_\chi ~\rra ~ \hc_\t    ~,\ee
\eqn{iaac} \htt^\t (g) \circ  {\tilde L}_\chi^\t
~=~ {\tilde L}_\chi^\t \circ  T^\chi (g) ~, \quad \forall g\in G ~,\ee
\eqn{inta} ( {\tilde L}_\chi^\t \ f ) (\tn_x a) ~=~
\int \kc (\tn_x,\va;\tn_{x'})\, f(\tn_{x'})\, d^dx' \ee
where ~$\kc (\tn_x,\va;\tn_x')$~ is a linear operator acting
from the space $V_\mu$ to the space $U_\t\,$,
and we have used the fact that ~$\tn_x \in \tN$ may be parametrized by $R^d$, 
 i.e., ~$x,x'\in R^d$, further $d^dx$ is the Haar measure on $R^d$. Actually, the integral kernel
depends only on the difference ~$z=x-x'$~ and is explicitly given by:
 \eqn{fff} \kc (z,\va) ~=~ N_\chi^\t\ \Big( {\va \over b(z) +\va^2}\Big)^{d-\D}\
\tD^\t (k_c(-{z\over \va} ) )\ \Pi^\mu_\t \ ,~~~ z=x-x',  \ee
where ~$b(z)$ is a bilinear form on $z\in \bbr^d$, ~$b(z)= z_1^2 + \cdots + z_d^2$,
and we use \eqref{kanman}, thus   $k_c$  from
 \eqref{dkk} is written explicitly as  ~$k_c(z)$.\\ 
 In \eqref{fff}   ~$N_\chi^\t$~ is arbitrary for
the moment and should be fixed from the requirement that ~${\tilde
L}_\chi^\t$~ is inverse to ~$L_\chi^\t\,$ (the latter being true except
on a parameter subspace of $(\chi,\t)$ of lower dimensionality).}~\dia

{\it Remark 1:} ~~The Theorem was proved in \cite{Dobads} for the
case ~$G=SO_0(p,1)$, with ~$d=p-1$. For the other real rank 1 cases
~$G=SU(r,1)$, ~$G=Sp(r,1)$, ~ $F_{4(-20)}$, the details will be
given in \cite{Dobpr}. In those cases the dimension of ~$\tN_0$~ is
~$2r-1$, $4r-1$, $15$, resp.\dia

{\it Remark 2:} ~~Note that the Theorem is valid also for the
replacement of the representation ~$\chi = [\mu,\nu]$~ by the
conjugate representation (called "shadow" in the physics
applications) ~$\tilde{\chi} = [\tilde{\mu},\tilde{\nu}]$, where
~$\tilde{\mu}$~ (called "mirror") is the Weyl conjugate of $\mu$,
while ~$\tilde{\nu}(a) = d-\D$.~\dia

 Furthermore, on the elementary representations $\chi$ are defined the integral Knapp-Stein ~$G_\chi$~
operators which intertwine the representation ~$\chi$~ with the representation ~$\tilde{\chi}$~:
\eqn{inter} G_\chi ~:~ C_{\tilde{\chi}} \ra C_{\chi} \ , ~~~
G_\chi \circ T_\chi (g) = T_{\tilde{\chi}} \circ G_\chi \ee
The operators $G_\chi$, $G_{\tilde{\chi}}$ have integral kernels that are given by the corresponding two-point functions, cf. \cite{DMPPT}, \cite{Dob1}.
The  representations  $\chi$, ${\tilde{\chi}}$ are
 called ~{\it partially equivalent}~   due to the existence of the above
  intertwining operators. The representations
are called ~{\it equivalent}~ if the latter intertwining operators
are onto and invertible.\\
We also recall that   the Casimirs
 $\cc_i$ have the same values on the partially   equivalent ERs:
\eqn{casiv} \cc_i(\mu,\nu) ~=~ \cc_i(\tilde{\mu}, \tilde{\nu}) \ee

Thus, a bulk representation has ~$two$~ elementary representations as boundaries!

\section{The split rank $>1$ cases}

\subsection{The split rank  two cases}

We restrict first the exposition to   the cases of split rank 2. This is natural since this includes the very important for applications Minkowskian conformal group
 ~$SO_0(p,2)$ in p-dimensions. (For p=4 see, e.g., \cite{DojmpMo}, for p=3 \cite{Dob32}.) 
 Furthermore, these cases are indicative for the general cases.
 Our results on the structure of the real Lie algebras of split rank 2 are given in Table 3.
For completeness besides the classical cases we have included also
 the  three exceptional real Lie algebra of split rank 2 ~:~  $E_{6(-14)}$, $E_{6(-26)}$ and  $G_{2(2)}$.

  For our considerations we shall use first the universal elementary representations introduced from a suitable maximal parabolic ~$P=MAN$~ so that
  the subgroup $N$ is of maximal dimension w.r.t. other possible choices.
For split rank 2 we have possibly two maximal parabolics shown as ~$M_1$ and $M_2$ in Table 3. Sometimes the two maximal parabolics  are isomorphic
and in that case there is only one entry for ~$M$~ in the table. We first consider the elementary representations induced from a maximal parabolic
~$P=MAN$, so that in  the case when two such parabolics are available we designate the chosen one again by   ~$M$.
Thus, we have:
\bigskip

\noindent\bu ~~{Bulk-boundary via Bruhat decomposition:}\\
This intertwining relation is similar to Theorem 1 above but   relations \eqref{aaa}, \eqref{aab}, \eqref{aac}
are replaced by:
\eqn{aaam}  L_\chi~:~  \check{C}_\chi ~\rra ~C_\chi \, , \ee
\eqn{aabb} (L_\chi\phi ) (\tn) ~=~ \lim_{\va\to 0}\
\va^{-\nu(a) }\   \phi (\tn a) \ee
\eqn{aacm} L_\chi  \circ \check{T}^\chi (g) ~=~
 T^\chi (g) \circ L_\chi \ ,\ee
while relations \eqref{prj},\eqref{prja} are not relevant as
 there is no factor  ~$\Pi^\t_\mu$.  \\
 Furthermore, relations \eqref{inv}, \eqref{iaac}, \eqref{inta}, \eqref{fff}
are replaced by:
 \eqn{invm} {\tilde L}_\chi
~:~ C_\chi ~\rra ~ \check{C}_\chi    ~,\ee
\eqn{iaacm} \check{T}^\chi   (g) \circ  {\tilde L}_\chi
~=~ {\tilde L}_\chi \circ  T^\chi (g) ~, \quad \forall g\in G ~,\ee
\eqn{intam} ( {\tilde L}_\chi \ f ) (\tn_x a) ~=~
\int K_\chi (\tn_x,\va;\tn_{x'})\, f(\tn_{x'})\, d^dx' \ee
 \eqnn{fffm} K_\chi  (z,\va) ~&=&~ N_\chi \Big( {\va \over b'(z) +\va^2}\Big)^{d-\D}\
D_\mu (r(-{z\over \va} ) )\   , \\
&&  ~~~  ~~~z=x-x',  ~~~ 
r(y) \in M,   \nn
\eea
where the explicit form of $b'(z)$ depends on the concrete $N$,
$ r(y)$ depends on the concrete $M$ and the concrete representation $\mu$
(see \cite{Dobpr} where some cases will be considered).

\bigskip

 \noindent\bu ~~{Bulk-boundary via Sekiguchi-like decomposition:}\\
 This intertwining relation is similar to Theorem 1 above. Actually, we just need to replace ~$K$ with $H$ - when it exists, and then replace $M_0$ with $M$
 (which is a subgroup of $H$). Thus, we have
 \eqn{aaah}  L_\chi^\s ~:~  \hc_\s ~\rra ~C_\chi \, , \ee
with the following action:
\eqn{aabh} (L_\chi^\s \phi ) (\tn) ~=~ \lim_{\va\to 0}\
\va^{-\D}\ \Pi^\s_\mu\ \phi (\tn a) \ee
where ~$\D=\nu(a)$, ~$\Pi^\s_\mu$~ is the standard projection operator from the
~$H$-representation space ~$W_\s$~ to the ~$M$-representation space
~$V_\mu\,$, which acts in the following way on the
$H$-representation matrices:
\eqn{prjh} \Pi^\s_\mu\ \tD^\s (H) ~=~ \hd^\mu (m(H))\   \Pi^\s_\mu \
\tD^\s (h_c) \ee
where we have used group decomposition for every $h\in H$~:
\eqn{dkkh} h ~=~ m(h) h_c\ee
where ~$m(h)$ parametrizes the subgroup $M\subset H$, while ~$h_c$~ parametrizes
the coset ~$H/M$, thus, \eqref{dkkh} represents the decomposition of
~$H$~ into its subgroup ~$M$~ and the coset ~$H/M$~: ~
~$H ~\cong ~ M~K/M$.  We shall use also the relation ~$H/M \cong \tN$~ following from:
\eqn{kanmanh} G= HAN \cong \tN MAN ~\Rightarrow ~H\cong \tN M ~\Rightarrow ~ H/M \cong \tN \ee
Then ~$L_\chi^\s$~ is an intertwining operator, i.e.:
\eqn{aach} L_\chi^\s   \circ  \htt^\s (g) ~=~
 T^\chi (g) \circ L_\chi^\s ~, \quad \forall g\in G ~.\ee
In addition, in \eqref{aabh}  the operator ~$\Pi^\s_\mu$~
acts in the following truncated way:
\eqn{prjah} \Pi^\s_\mu\ \tD^\s (h) ~=~ \hd^\mu (m(h))\   \Pi^\s_\mu \ee
  The operator inverse to ~$L_\chi^\s\,$~ which
would restore a function on de Sitter space $\cs$ from its boundary
value is given as   follows:
\eqn{invh} {\tilde L}_\chi^\s
~:~ C_\chi ~\rra ~ \hc_\s   ~,\ee
\eqn{iaach} \htt^\s (g) \circ  {\tilde L}_\chi^\s
~=~ {\tilde L}_\chi^\s \circ  T^\chi (g) ~, \quad \forall g\in G ~,\ee
\eqn{intah} ( {\tilde L}_\chi^\s \ f ) (\tn_x a) ~=~
\int K^\s_\chi  (\tn_x,\va;\tn_{x'})\, f(\tn_{x'})\, d^dx' \ee
where ~$K^\s_\chi (\tn_x,\va;\tn_x')$~ is a linear operator acting
from the space $V_\mu$ to the space $W_\s\,$,
and we have used the fact that ~$\tn_x \in \tN$ may be parametrized by $R^d$, 
 i.e., ~$x,x'\in R^d$, further $d^dx$ is the Haar measure on $R^d$. Actually, the integral kernel
depends only on the difference ~$z=x-x'$~ and is  given by:
 \eqn{fffh} K^s_\chi (z,\va) ~ =~ N_\chi^\s\ \Big( {\va \over b''(z) +\va^2}\Big)^{d-\D}\
\tD^\s (h_c(-{z\over \va} ) )\ \Pi^\mu_\s \  .  \ee
where the explicit form of $b''(z)$ depends on the concrete $N$,
$ h_c(y) $ depends on the concrete $H$ and the concrete representation $\s$.

\md
\noindent {\it Remark:} ~Sekiguchi \cite{Seki} introduced this decomposition for the cases AIII, BDI, CII, though not in our context. In the AdS/CFT context
in the case BDI for $G=SO_0(d,2)$ the local coordinates of  $\tN A$ are \cite{DoMo2}:
\eqnn{sekigus}
 && x_\mu =   \frac{g_{\mu,d+1}} { g_{d+1,d} + g_{d+1,d+1}}\ , ~~\mu=0,\ldots,d-1 ,\nn\\
 && \va =   | g_{d+1,d} + g_{d+1,d+1} |  ,\eea
where the matrix
 $g\in G$ is represented explicitly by ~$g_{\a\b}$, ~$\a,\b=0,1,\ldots,d, d+1$.\dia

\subsection{Split rank $>2$ cases}

Our results on the structure of the cases of higher split rank $>2$ are given in Tables 4 for   classical real semisimple Lie groups
and in Table 5 for  exceptional  real semisimple Lie groups and algebras.

In Table 4 we give the important factors when using maximal parabolics for classical real semisimple Lie groups. The various parabolics are enumerated
by giving explicitly the factors $M_j$ in column 4.  We also give the Sekiguchi-like factors $H$ when available.
In some cases $H$ coincides with some $M_j$ factor and this is pointed out in Column 3. In other cases it is important
to record which $H$ factor is consistent with some factor $M_j$ ~: ~$M_H=M_j$, so that the decompositions hold:
\eqn{parahh}  G ~\cong~ \tN A_m H ~\cong~  \tN A_m M_H N \ee

In Table 5 we give the important factors when using maximal parabolics exceptional real semisimple Lie algebras.
In the case there is no parametric enumeration of the ~$\cm_j$~ factors, so the possible cases are given explicitly.
Here there are less occurrences of Sekiguchi-like factors ~$\ch_j$~ and they are given next to the
consistent with them ~$\cm_j$~ factors:
\eqn{parahhz}  \cg ~\cong~ \tilde{\cn}  \ca_m \ch ~\cong~  \tilde{\cn}  \ca_m \cm_\ch \cn \ee

The bulk-boundary correspondence is given similarly to the Split rank 2 cases. In the exceptional cases we use the language
of Lie algebras. This could be important for the subtle differences between the Lie groups $M,H$ and their Lie algebras $\cm,\ch$,
yet there is no problem since we use finite-dimensional representations of $M$ and $H$ for the induction process.

\section{Summary and Outlook}

 In the present paper we have laid down the foundations of generalizing the AdS/CFT (holography) idea   beyond the conformal setting.
The main tool is to find suitable realizations of the bulk and boundary via group theory.  We use all ten families of classical real semisimple Lie groups $G$
and Lie algebras $\cg$. On the boundaries we use the notion of elementary representations since these provide as subrepresentations all possible irreducible 
admissible representations of $G$. In the bulk we use several group and algebra decompositions choosing what is most simple to use. Thus, we use 
the global Iwasawa decomposition in the cases when $G$ is of split rank one. In the cases when $G$ is of split rank $>1$ we use the local 
Sekiguchi-like decomposititions when it exists, otherwise we use the local Bruhat decomposition. All results are given in separate tables. 
 The same analysis is applied to the exceptional real semisimple Lie algebras.

We stress that these are only the foundations. Further investigations for explicit applications would require separate work on each member of the 
ten families of classical real semisimple Lie groups, also separately for the cases of split rank one and split rank $>1$, also taking into account 
the availability or not of the Sekiguchi-like decomposititions. 

Among further more remote possible applications we would mention the following.
 One may look to accommodate in some cases of our setting a time direction and energy associated with it, 
 maybe using the fact that in some cases there exist positive energy representations via holomorphic highest weight representations. 
 Certainly, similar ideas from the present paper may be applied to more general symmetry objects such as:  quantum groups, supergroups, Kac-Moody groups, 
 especially, if suitable for applications to string theory.


\vspace{10mm}

\section*{Acknowledgments}

The author thanks the reviewer for numerous remarks that helped in improving the exposition. 
The author has received partial support from  Bulgarian NSF Grant DN-18/1.

\vspace{10mm}

\newpage

\parskip=4pt
\baselineskip=12pt
\parindent 10pt
\voffset .5cm

\def\hann{{\textstyle\frac{n}{2}}}

\def\tablerule{\noalign{\hrule}}

\pagestyle{empty}

\hskip-2cm\vbox{\centerline{\bf   Table 1 : Tenfold list of}\medskip \hfil\break 
\centerline{\bf classical real semisimple Lie groups $G$}
\smallskip
\offinterlineskip \halign{\baselineskip12pt
\strut\vrule#\hskip0.1truecm &
#\hfil& \vrule#\hskip0.1truecm &
#\hfil& \vrule#\hskip0.1truecm  &
#\hfil& \vrule#\hskip0.1truecm  &
#\hfil& \vrule#\hskip0.1truecm  &
#\hfil& \vrule#\hskip0.1truecm  &
#\hfil& \hskip0.1truecm  \vrule#
\cr \tablerule &&&&& &&&&& &&\cr
&Type  && $G=KA_0N_0$ &&  $K$maximal&&$\dim A_0N_0$ &&split& & $M_0$&\cr
&of $G$&&  $\dim G$&& compact subgroup&&&&rank&& &\cr \tablerule &&&&&& &&&&&&\cr
& A  && $SL(n,\bbc)_\bbr$ && $SU(n)$ && $n^2-1$&& $n-1$ &&$U(1)\times ... \times U(1)$&\cr
 &&& $2(n^2-1)$&& &&&&&&$n-1$ times&\cr \tablerule &&&&& &&&&&& &\cr
& AI &&$SL(n,\bbr)$
 && $SO(n)$     && $\ha (n^2+n-2)$ &&$n-1$&&{\bf e}&\cr
  &&&$n^2-1$&&&&&&&&&\cr
 \tablerule &&&&&&&&&&&&\cr
&AII &&  $SU^*(2n)$  && $Sp(n)$&& $2n^2-2n+1$&&$n-1$&& $SU(2)\times ... \times SU(2)$&\cr
&&&$4n^2-1$&&&&&&&&$n$ times&\cr \tablerule &           &&&&&&&&&& $SU(p-q)\times U(1)\times ... \times U(1)$&\cr
&AIII &&  $SU(p,q)$ && $S(U(p)\times U(q))$ &&$2pq$&& $q$&&$p>q$, \hskip 2cm $q$ times&\cr
&&&$(p+q)^2-1$  &&&&&&&&  $U(1)\times ... \times U(1)$, $q-1$ times&\cr
&&&$p\geq q$  &&&&&&&&$p=q$   &\cr
 \tablerule &&&&&&&&&&&&\cr
&BD &&$SO(n,\bbc)_\bbr$ &&$SO(n)$ && $\ha n(n-1)$&& $[\hann]$&&$U(1)\times ... \times U(1)$&\cr
&&&$n(n-1)$&&&&&&&&$[\hann]$ times&\cr &&&$n>2$&&&&&&&&&\cr\tablerule &&&&&&&&&&&&\cr
&BDI &&$SO_0(p,q)$ &&$SO(p)\times SO(q)$ &&$pq $&& $q$&&$SO(p-q)$&\cr
&&&$(p+q)(p+q-1)/2$  &&&&&&&&&\cr
&&&$p\geq q$  &&&&&&&&&\cr
\tablerule&&&&&&&&&&&$SO(3)\times ...\times SO(3)$&\cr
&DIII  && $SO^*(2n)$  && $U(n)$&& $n(n-1)$ &&$[\hann]$&&  $n=2r$, $r$ times&\cr
&&&$n(2n-1)$&& &&&&&&$SO(2)\times SO(3)\times ...\times SO(3)$&\cr &&&&&&&&&&&  $n=2r+1$, $r$ times&\cr\tablerule &&&&&&&&&&&&\cr
&C && $Sp(n,\bbc)_\bbr$  && $Sp(n)$&& $n(2n+1)$&& $n$&&$U(1)\times ... \times U(1)$ &\cr
&&&$2n(2n+1)$&& &&&&&&$n$ times&\cr \tablerule &&&&&&&&&&&&\cr
&CI && $Sp(n,\bbr)$  && $U(n)$&& $n(n+1)$&&$n$&& {\bf e}&\cr
 &&& $n(2n+1)$&&&&&&&&&\cr\tablerule &&&&&&&&&&&&\cr
&CII && $Sp(p,q)$&& $Sp(p)\oplus Sp(q)$   &&$4pq$ &&$q$&&$sp(p-q)\times sp(1)\times ... \times sp(1)$&\cr
&&&$(p+q)(2(p+q)+1)$   && &&&&&& \hskip 3cm $q$ times&\cr
&&&$p\geq q$  && &&&&&&&\cr
\tablerule
 }}

\vbox{\hskip3truecm{\bf Table 2 : Tenfold list of}\medskip \hfil\break
{\bf   real semisimple Lie groups $G$ of split rank 1 ($= \dim A_0$) },\medskip\\
{\bf  some represented by the Euclidean conformal group $SO_0(p,1)$}
\smallskip
\offinterlineskip \halign{\baselineskip12pt
\strut\vrule#\hskip0.1truecm &
#\hfil& \vrule#\hskip0.1truecm &
#\hfil& \vrule#\hskip0.1truecm  &
#\hfil& \vrule#\hskip0.1truecm  &
#\hfil& \vrule#\hskip0.1truecm  &
#\hfil& \hskip0.1truecm  \vrule#
\cr \tablerule &&&&& &&&&&\cr
&Type  && $G=KA_0N_0$ &&  $K$maximal&&$\dim A_0N_0$&& $M_0$& \cr
&of $G$&&$\cong \tN_0M_0A_0N_0$&& compact subgroup&&bulk dim.&&&\cr \tablerule &&&&&&&&&&\cr
&A$\cong$BDI && $SL(2,\bbc)_\bbr$ && $SU(2)$ && $3$&& $U(1)$&\cr
&$p=3$  && && &&&&&\cr
\tablerule &&&&&&&&&&\cr
&AI$\cong$BDI &&$SL(2,\bbr)$ && $SO(2)$     && $2$ &&${\bf e}$&\cr
  &$p=2$&&&&&&&&&\cr
\tablerule &&&&& &&&&&\cr
&AII$\cong$BDI &&  $SU^*(4)$  && $Sp(2)$&& $5$&&$SU(2)\times SU(2)$&\cr
&$p=5$&&&& &&&&&\cr \tablerule &&&&& &&&&&\cr
&AIII &&  $SU(r,1)$ && $U(r)$ &&$2r$&& $U(r-1)$&\cr
&&&$r\geq 2$  &&&&&&&\cr
 \tablerule &&&&&&&&&&\cr
&BD $\cong$ BDI  &&$SO(3,\bbc)_\bbr$ &&$SO(3)$ && $3$&& $SO(2)$ &\cr
&  $p=3$&&&&&&&&&\cr\tablerule
&&& &&& &&&&\cr
&BDI  &&$SO_0(p,1)$ &&$SO(p)$ &&$p$&&  $SO(p-1)$  &\cr
&&&$p\geq 2$  &&&&&&&\cr
\tablerule&&&&&&&&&&\cr
&DIII  && $SO^*(4)$  && $SO(3)\times SO(2)$ && $2$ &&$SO(3)$ &\cr
&&&$ =SO(3)\times SO_0(2,1)$&& &&&& 
&\cr \tablerule &&&&&&&&&&\cr
&C$\cong$BDI && $Sp(1,\bbc)_\bbr$  && $Sp(1)$&& $3$&& $U(1)$&\cr
&$p=3$&&&& &&&&&\cr \tablerule &&&&&&&&&&\cr
&CI$\cong$BDI && $Sp(1,\bbr)$  && $U(1)$&& $2$&&${\bf e}$&\cr
&$p=2$&&&& &&&&&\cr\tablerule &&&&& &&&&&\cr
&CII && $Sp(r,1)$&& $Sp(r)\oplus Sp(1)$   &&$4r$ &&$Sp(r-1)\oplus Sp(1)$  &\cr
&&&$r\geq 2$  &&&&& &&\cr
\tablerule
&&& &&& &&&&\cr
&FII && $F_{4(-20)}= F''_4$     && $so(9)$&& $16$&&$so(7)$ &\cr
&&& &&& &&&&\cr\tablerule
 }}

\np

\small

\hskip-2truecm\vbox{\hskip3truecm{\bf Table 3 : Tenfold Table of}\medskip \hfil\break 
{\bf   real semisimple Lie groups $G$ of split rank 2, ($\dim A_0=2$),}\medskip\\
{\bf    showing maximal parabolic subgroups $P=MA_mN$, ($\dim  A_m=1$)},\medskip\\
{\bf showing also Sekiguchi(-like) subgroups $H$}\\
\smallskip
\offinterlineskip \halign{\baselineskip12pt
\strut\vrule#\hskip0.1truecm &
#\hfil& \vrule#\hskip0.1truecm &
#\hfil& \vrule#\hskip0.1truecm  &
#\hfil& \vrule#\hskip0.1truecm  &
#\hfil& \vrule#\hskip0.1truecm  &
#\hfil& \hskip0.1truecm  \vrule#
\cr \tablerule
&&& $G=KA_0N_0$ && &&&&$\dim  \tN A_m$&\cr
&Type  && $\cong \tN A_mMN$ &&~$K$    &&$M$   &&$=\dim G/MN$ &\cr
&of $G$&&$\cong \tN A_mH$ &&~$M_0$ && $H$  && $(=\dim G/H)$&\cr
&&& $\dim G$&&&&&& = bulk dim. &\cr
&&&&&&&&&= d+1&\cr
\tablerule
& A  && $SL(3,\bbc)_\bbr$ && $SU(3)$ &&$M=U(1)\times SL(2,\bbc)_\bbr$ && $5$&\cr
 &&& $16$ &&  $U(1)\times U(1)$&& &&&\cr \tablerule 
& AI &&$SL(3,\bbr)$
 && $SO(3)$     && $M=SL(2,\bbr)$ &&$3$&\cr
  &&&$8$&&{\bf e}&&  &&&\cr\tablerule 
&AII &&  $SU^*(6)$  && $Sp(3)$&& $M=SU(2)\times SU^*(4)$ &&$9$&\cr
&&&$35$&& $SU(2)\times SU(2)\times SU(2)$&&  &&&\cr \tablerule 
&AIII &&  $SU(p,2)$ && $  U(p)\times SU(2)$ &&$ M_1=M=U(p-1,1)$ &&$2(p+1)$&\cr
&&&$p^2+4p+3$  &&$  U(p-2)\times U(1)$&&  $ M_2=U(p-2)\times SL(2,\bbc)_\bbr$ && &\cr
&&&$p\geq 2$  &&&&$ H=U(p,1)$ &&&\cr
 \tablerule &&&&&&&$M_1=M=SO(2)\times SO(2,\bbc)_\bbr$&&&\cr
&BD &&$SO(4,\bbc)_\bbr$ &&$SO(4)$ && $M_2=SO(2)\times SL(2,\bbc)_\bbr$ &&$5$&\cr
&&&$12$&& $SO(2)\times SO(2)$ &&$H=SO(2)\times SO(3,\bbc)_\bbr$&&&\cr
\tablerule 
&BD &&$SO(5,\bbc)_\bbr$ &&$SO(5)$ &&  $M=SO(2)\times SO(3,\bbc)_\bbr$ &&$7$&\cr
&&&$20$&&$SO(2)\times SO(2)$&&$H=SO(2)\times SO(4,\bbc)_\bbr$&&&\cr
\tablerule
&BDI &&$SO_0(p,2)$ &&$SO(p)\times SO(2)$ &&$M_1=M=SO_0(p-1,1)$ &&$p+1$&\cr
&&&$(p+2)(p+1)/2$  &&$SO(p-2)$ && $M_2=SO(p-2)\times SL(2,\bbr)$ &&&\cr
&&&$p\geq 2$  &&&&$H=SO_0(p,1)$&&&\cr
\tablerule
&&&&&&&$M_1=M=SO(3)\times SO^*(4)$&&&\cr
&DIII  && $SO^*(8)$  && $U(4)$&&$M_2=SO_0(5,1) $  &&$10$&\cr
&&&$28$&&$SO(3) \times SO(3)$ &&$H=SO(3)\times SO^*(6)$ &&&\cr \tablerule
&&&&&  $U(5)$ && $M_1=M=SO(3)\times SO^*(6)$&&&\cr
&DIII  && $SO^*(10)$  &&  $SO(2) \times SO(3)\times $&&
$M_2=SO_0(5,1)\times SO(2)$  &&$14$&\cr
&&&$45$&&  $\times SO(3)$&&$H=SO(3)\times SO^*(8)$  &&&\cr \tablerule
&C && $Sp(2,\bbc)_\bbr$  &&$Sp(2)$ && $M=U(1)\times Sp(1,\bbc)_\bbr$    &&$7$&\cr
&&&$20$&& $U(1)\times U(1)$ && $ H= U(1)\times Sp(1,\bbc)_\bbr\times$ &&&\cr
&&&&&&&$\times Sp(1,\bbc)_\bbr$&&&\cr
\tablerule 
&CI && $Sp(2,\bbr)$  && $U(2)$&&$M=Sp(1,\bbr)$  &&$4$&\cr
 &&&$10$&&{\bf e}& &$H=Sp(1,\bbr)\times Sp(1,\bbr)$&&&\cr\tablerule 
&CII && $Sp(p,2)$&& $Sp(p)\times Sp(2)$   && $M_1=M=Sp(1)\times
Sp(p-1,1)$ &&$4(p+1)$&\cr
&&&$(p+2)(2p+5)$  &&$Sp(p-2)\times Sp(1)\times$&& $M_2=Sp(p-2)\times SU^*(4)$  &&&\cr
&&&$p\geq 2$  && $\times Sp(1)$ &&$H=Sp(1)\times Sp(p,1)$&&&\cr
  \tablerule 
 &EIII  && $E_{6(-14)}=E^{iii}_6$  && $so(10)\oplus so(2)$&&
$\cm_1=so(7,1)\oplus so(2)$&&$25$&\cr
&&&78 && $so(6)\oplus so(2)$& & $\cm_2=su(5,1)$ & & 22&\cr
 \tablerule 
 &EIV   && $E_{6(-26)}=E^{iv}_6$  && $f_4$&&
$\cm =so(9,1)$ &&$17$&\cr
&&&78 && $so(8)$& & 
& &&\cr\tablerule
&G && $G_{2(2)}= G'_2$   && $so(3)\oplus so(3)$&& $\cm=sl(2,\bbr)$&&  6&\cr
  &&& 14&&{\bf e}& & $\ch=sl(3,\bbr)$&& &\cr
\tablerule
 }}

\np

\hskip-3.5truecm\vbox{\hskip3truecm{\bf Table 4 : Tenfold Table of}\medskip \hfil\break 
{\bf classical real semisimple Lie groups ~$G$~ showing}\medskip\\
{\bf maximal parabolic subgroups ~$P=MA_mN$, ~$\dim A_m=1$, }\medskip\\
{\bf also Sekiguchi(-like) subgroups}
\smallskip
\offinterlineskip \halign{\baselineskip12pt
\strut\vrule#\hskip0.1truecm &
#\hfil& \vrule#\hskip0.1truecm &
#\hfil& \vrule#\hskip0.1truecm  &
#\hfil& \vrule#\hskip0.1truecm  &
#\hfil& \vrule#\hskip0.1truecm  &
#\hfil& \hskip0.1truecm  \vrule#
\cr \tablerule
&Type  && $G  \cong \tN  A_mH$ &&~$H$~Sekiguchi(-like)  &&~$M_j$   &&~bulk dim.:  &\cr
&of $G$&&$\cong \tN A_mMN$ &&~~~~subgroup&&  &&~$\dim \tN_j A_m=d+1=$&\cr
&&& $\dim G$ &&&&&& $= \dim G/H $ &\cr
\tablerule
& A  && $SL(n,\bbc)_\bbr$ && $n=7$ $\downarrow$ $H\cong M_1$    &&$U(1)\!\times\! SL(j,\bbc)_\bbr\!\times\!   SL(n-j,\bbc)_\bbr$ && $2j(n-j)+1$ &\cr
 &&& $2(n^2-1)$ && $H = U(1)\times SL(6,\bbc)_\bbr$ && $\dim M_j = 2 (n^2 +2j^2 -2nj  ) -3$ && $n=7, j=3,4$$\downarrow$ &\cr
 &&&&& $\dim H = 71$&&
 &&$\dim \tN_{3,4} A_m=25$&\cr \tablerule
& AI &&$SL(n,\bbr)$
 &&  $n=7$ $\downarrow$ $H\cong M_1$   && $SL(j,\bbr)\oplus SL(n-j,\bbr)$ &&$j(n-j)+1$&\cr
  &&&   $ n^2-1 $  &&$H= SL(6,\bbr)$&& $\dim M_j = n^2 +2j^2 -2nj   -2$  && $n=7, j=3,4$ $\downarrow$&\cr
  &&&&& $\dim H = 35$&& && $\dim \tN_{3,4} A_m=13$&\cr\tablerule
&AII &&  $SU^*(2n)$  &&   && $SU^*(2j)\times SU^*(2n-2j)$ &&$
4j(n-j)+1$&\cr &&&$ 4n^2-1 $&&&&$\dim M_j= 4(n^2-2nj+2j^2)-2$ &&
&\cr  
\tablerule &
&&&& &&$M_j\!=\! U(p-j,q-j)\!\times\! SL(j,\bbc)_\bbr\! $  &&&\cr
&AIII &&
$SU(p,q)$ && $U(p,q-1)$  &&  for ~$j\leq q$, ~$M_H=M_1$&&j(2(p+q)-3j)+1&\cr
&&&$(p+q)^2-1$ &&$p\geq q>1$&& $n^2 +6j^2-4nj-2$, n=p+q   &&&\cr
 \tablerule &&& &&&&  $M_j=SO(2) \times SL(j,\bbc)_\bbr\times $  &&&\cr
&BD &&$SO(n,\bbc)_\bbr$ && $SO(2)\times SO(n-1,\bbc)_\bbr$ &&$\times  SO(n-2j,\bbc)_\bbr$, $j\!\leq\![{n\over 2}]$
&&$2jn-3j^2-j+1$&\cr
&&& $n(n-1), n>2$ && $n^2-3n+3$ && $n^2+6j^2-4jn-n+2j-1$ &&&\cr
&&&& &&& $M_H=M_1$ &&  &\cr
\tablerule
&&&&&&& $M_j=SL(j,\bbr)\times SO_0(p-j,q-j)$&&&\cr
&BDI &&$SO_0(p,q)$ && $SO_0(p,q-1)$ && j$\leq q<p$ or j$\leq$ q-4, for p=q$\geq$5&&$j(p+q) -\textstyle{3j^2+j\over 2}+1$  &\cr
&&&$(p+q)(p+q-1)/2$&&$p> q>1$, $p=q>2$&& $\textstyle{n^2+6j^2-4jn-n+2j-2\over 2}$, n=p+q&&&\cr
&&& && ~$M_H=M_1$ &&$M_q\!=\!M_{q-1}\!=\! SL(q,\bbr)$ for p=q$\geq$2 &&$\textstyle{q(q-1)\over 2}$+1 &\cr
&&& &&&&$M_{q-2}= SL(q-2,\bbr)\times  $   &&&\cr
 &&&&&$M_{q-3}= SL(q-3,\bbr)\times$ && $\times\!SL(2,\bbr)\!\times\!SL(2,\bbr)$ for p=q$\geq$3 &&$\textstyle{q^2+3q\over 2} -4$&\cr
  &&&&&$\times SL(4,\bbr)$ for p=q$\geq$4 && $q^2-4q+9$&& &\cr
  &&&&&$q^2-6q+23~~\ra \ra\ra$ &&&&$\textstyle{q^2+5q\over 2} -11$&\cr
\tablerule  &DIII  && $SO^*(2n)$  && $SO^*(2n-2)\times SU(2)$&&
$M_j = SO^*(2n-4j) \times SU^*(2j)$
&&$4j(n-j)-$&\cr &&& $n(2n-1)$&& &&$j\leq [\hann]$, ~$M_H=M_1$  &&$-2j^2-j+1$ &\cr
\tablerule
&C &&$Sp(n,\bbc)_\bbr$ && U(1)$ \times  Sp(1,\bbc)_\bbr  \times\!$
 && $U(1)\!\times\! SL(j,\bbc)_\bbr \!\times\!
Sp(n-j,\bbc)_\bbr$ &&$4nj+j-3j^2 +1$ &\cr
 & && $2n(2n+1)$  && $\times Sp(n-1,\bbc)_\bbr$&& $ 4n(n-2j)+6j^2+2(n-j)-1$ &&  $\dim \tN_2A_m$ = 8n-9&\cr
 &&& &&$\dim H = 4n^2 -6n +9$ && $M_H=M_2$ && &\cr 
 \tablerule
 &  CI &&$Sp(n,\bbr)$   && $Sp(1,\bbr)\times Sp(n-1,\bbr) $ && $M_j=SL(j,\bbr)\times Sp(n-j,\bbr)$&& $2nj + {j-3j^2 \over 2}+ 1$&\cr
& && $n(2n+1)$ && $\dim H = 2n^2 -3n +4$ && $ 2(n^2-2nj)+3j^2+n-j-1$  && $\dim \tN_2A_m$= 4n-4 &\cr
 &&& &&    &&  $M_H=M_2$  &&&\cr 
 \tablerule &&&&&&& $Sp(p-j,q-j)\times SU^*(2j),\ j\leq q$&&&\cr
&CII && $Sp(p,q)$&& $Sp(1)\times
Sp(p,q-1)$    &&$ 2(n^2-4nj +6j^2)+n-2j-1$ &&$j(4(p+q)-6j)+$&\cr
&&&$(p+q)(2(p+q)+1) $  &&$p\geq q>1$ &&  ~n=p+q, ~ $M_H=M_1$     &&$+j+1$&\cr
 \tablerule
 }}

\np

\hskip-3.5cm\vbox{\centerline{\bf   Table 5}\medskip \hfil\break 
{\bf exceptional real semisimple Lie groups $G$, resp. algebras $\cg$,\\ of split rank $>$ 2}
\smallskip
\offinterlineskip \halign{\baselineskip12pt
\strut\vrule#\hskip0.1truecm &
#\hfil& \vrule#\hskip0.1truecm &
#\hfil& \vrule#\hskip0.1truecm  &
#\hfil& \vrule#\hskip0.1truecm  &
#\hfil& \vrule#\hskip0.1truecm  &
#\hfil& \vrule#\hskip0.1truecm  &
#\hfil& \hskip0.1truecm  \vrule#
\cr \tablerule
&Type  && $G=KA_0N_0$ &&  $\ck$ &&$\dim \tilde{\cn}_0 \ca_0$&& $\cm$ & &  $\dim \tilde{\cn} \ca_m$ =&\cr
& && $\cg=\ck\oplus\ca_0\oplus\cn_0$ &&  $\cm_0$ && split rank&& $\ch$&& = d+1 &\cr
& && $\cg=\tilde{\cn} \oplus \ca_m \oplus\cm\oplus\cn$ &&& &&& & &&\cr
  \tablerule
&&&&& & &&& $\cm_1=so(5,5)$ &&17 &\cr
 &EI &&$E_{6(6)}=E'_6$  && $sp(4)$&& $42$&&$\cm_2=sl(5,\bbr) \oplus sl(2,\bbr)$  && 26 &\cr
   &&&&& {\bf e} &&6 &&$\cm_3=sl(3,\bbr)\oplus sl(3,\bbr) \oplus sl(2,\bbr)$  &&30 &\cr
  &&&&&&&&& $\ch_3=so(5,5) \oplus sl(2,\bbr)$ && 30&\cr
    &&&&&&&&& $\cm_4=sl(6,\bbr)$ &&22 &\cr
   \tablerule
  &&&&&&&&&$\cm_1=so(5,3) \oplus so(2)$  && 25 &\cr
  &EII  &&$E_{6(2)}=E''_6$ && $su(6)\oplus su(2)$&& $40$&& $\cm_2=sl(3,\bbr) \oplus u(1)\oplus sl(2,\bbc)_\bbr$   &&32  &\cr
  &&&&& $u(1)\oplus u(1)$&& 4&&$\cm_3=sl(2,\bbr)\oplus sl(3,\bbc)_\bbr$ && 30 &\cr
  &&&&&&&&&$\cm_4=su(3,3)$ &&22 &\cr
  \tablerule
  &&&&& &&&& $\cm _1   = so(6,6)$  & &34 &\cr
 &EV   && $E_{7(7)}=E'_7$  && $su(8)$&& $70$&&  $\cm _2 = sl(6,\bbr) \oplus sl(2,\bbr)$  & &48 &\cr
 &&&&&{\bf e} &&7&& $\cm_3=sl(4,\bbr)\oplus sl(3,\bbr)\oplus sl(2,\bbr)$ && 54 &\cr
&&&&& &&&& $\ch_3 = E'_6\oplus so(2)$&&54&\cr
&&&&& &&&& $\cm_4 = sl(5,\bbr) \oplus sl(3,\bbr)$ & &51 &\cr
&&&&& &&&& $\cm _5   = so(5,5) \oplus sl(2,\bbr)$ & &43 &\cr
&&&&& && && $\cm_6=E'_6$ &&28 &\cr
&&&&& &&&& $\cm_7 = sl(7,\bbr)$  & &43 &\cr
\tablerule
&&&&&&&&& $\cm_1=so(7,3) \oplus su(2)$ && 43 &\cr
&EVI   &&$E_{7(-5)}=E''_7$  && $so(12)\oplus so(3)$ && $64$&& $\cm_2= sl(3,\bbr) \oplus su^*(4) \oplus su(2)$   &&54 &\cr
&&&&& $so(3)\oplus so(3)\oplus\quad$&&4 &&  $\cm_3 = sl(2,\bbr)\oplus  su^*(6)$  &&48 &\cr
&&&&&$\,\oplus so(3)$ && && $\cm_4 = so^*(12)$  &&34 &\cr
\tablerule
&&&&&&&&& $\cm_1=e^{iv}_6$ && 28 &\cr
 &EVII   && $E_{7(-25)}=E'''_7$  && $e_6\oplus so(2)$&& $54$&& $\cm_2=sl(2,\bbr) \oplus so(9,1)$  &&43   &\cr
 &&&&& $so(8)$&& 3&& $\cm_3=so(10,2)$  && 34 &\cr
 \tablerule
&&&&&&&&& $\cm_1 = so(7,7)$  && 79 &\cr
&EVIII   &&  $E_{8(8)}= E'_8$ && $so(16)$&& $128$&&$\cm_2 =  sl(7,\bbr) \oplus sl(2,\bbr)$  && 99 &\cr
&&&&&{\bf e} & & 8&&  $\cm_3 =  sl(5,\bbr) \oplus sl(3,\bbr) \oplus sl(2,\bbr)$ &&107 &\cr
 &&&&& && && $\cm_4 =  sl(5,\bbr) \oplus sl(4,\bbr)$ &&105 &\cr
 &&&&&&&&& $\cm_5   =  so(5,5) \oplus sl(3,\bbr)$ &&98 &\cr
&&&&&&&&& $\cm_6 =  E'_6 \oplus sl(2,\bbr)$ &&84 &\cr
&&&&&&&&& $\cm_7 =   E'_7$  &&58 &\cr
&&&&&&&&& $\cm_8 =  sl(8,\bbr)$ &&93 &\cr
\tablerule
&&&&&&&&& $\cm_1=so(11,3)$ && 79 &\cr
&EIX   &&  $E_{8(-24)}= E''_8$ && $e_7\oplus so(3)$&& $112$&& $\cm_2=sl(3,\bbr) \oplus so(9,1) $ && 98 &\cr
&&&&& $so(8)$ && 4&&  $\cm_3= sl(2,\bbr)\oplus e^{iv}_6$ && 84   &\cr
&&&&& && && $\cm_4=e'''_7$ && 58 &\cr
\tablerule 
&&&&&   &&&& $\cm_1=sl(3,\bbr) \oplus sl(2,\bbr) $  &&21 &\cr
&FI   && $F_{4(4)}=F'_4$  && $sp(3)\oplus so(3)$&& $28$&& $\cm_2=so(4,3)$  &&16  &\cr
&&&&& {\bf e}&&4&& $\ch_2=so(5,4)$  && 16&\cr
&&&&&   &&&& $\cm_3= sp(3,\bbr)$  &&16 &\cr
&&&&&   &&&& $\ch_3= sp(4,\bbr)$ &&16&\cr
 \tablerule }  }

\end{document}